# Experimental Study to Optimise the Treatment Efficacy of Pharmaceutical Effluents by Combining Electron Beam Irradiation with Conventional Techniques

Pankaj Kumar, Manisha Meena, Anjali Bhagwan Kavar,
Pragya Nama, Abhishek Pathak, Raghava Varma

Department of Physics
Indian Institute of Technology Bombay
Powai, Mumbai, 400076, India

Abhay Deshpande, Tanuja Dixit, R. Krishnan, Chandrakant Nainwad

Society for Applied Microwave Electronics Engineering & Research (SAMEER)
R&D Lab of Govt of India
Mumbai, 400076, India

## Abstract

The inability of conventional methods to completely remove the contaminants from pharmaceutical effluents led us to study the effect of Electron Beam (EB) irradiation on real pharmaceutical wastewater. In this paper, the samples from different stages of existing treatment facilities of industry are irradiated with varying doses from 25 to 200 kGy. The study aimed to find a suitable combination of EB and conventional treatments for efficient degradation of complex pharmaceutical effluent. It has been successfully demonstrated that electron beam irradiation when combined with conventional techniques like coagulation before or after the irradiation improves the efficiency of the process, resulting in lower Chemical Oxygen Demand (COD). In this investigation, the maximum COD reduction was found to be around 65 percent.

*Corresponding author: Pankaj kumar; E-mail: pankaj.r@iitb.ac.in;

## 1 Introduction

The water crisis across the world is now a reality. Climatic changes are not the only reason for this crisis, but a great part of it is primarily due to the non-sustainable use of our natural resources. To make matters worse, the population growth with rapidly growing standards of living have further introduced new pollutants called Contaminants of Emerging Concern (CECs) that include micro-pollutants, endocrine disruptors (EDs), pesticides, pharmaceuticals, hormones, toxins, and synthetic dyes ((Rasheed et al. (2019))). Amongst them, the pharmaceutical effluents are the most challenging to treat by conventional methods due to the presence of ammonium nitrogen, toxic and complex compounds that are produced as by-products in the production of anti-hypertensive, anti-asthmatic, anti-cancer, skin-care drugs. The types of impurities in pharmaceutical effluent depend on whether organic, inorganic or biological reactions are used in their production.

The production of drugs or pharmaceutical ingredients consists of 4 major processes, Chemical synthesis, Fermentation, Biological extraction and Formulation (Gadipelly et al. (2014)). In chemical synthesis, large amounts of intermediate by-products including acids, bases, halides, nitrates, sulphates, cyanides, and metals are produced. In addition, organic solvents like benzene, phenol, toluene and cyanide are also used in this process. All of them finally end up in the effluent making it difficult to treat. The wastewater from this process has a high amount of chemical oxygen demand (COD), biological oxygen demand (BOD) and total suspended solids (TSS). Fermentation process further increases the COD and BOD of the wastewater as it involves disinfectants, phenols and detergents. Hexane is most often used as a solvent for herbal extraction in biological extraction processes. The extracted chemicals with pharmaceutical ingredients and disinfectants find their way into the main waste stream from the biological extraction process. After these three processes, the drug products are converted into usable form like tablets and syrups. This process is known as the formulation process in which different types of chemicals like flavouring agents, preservatives, and antioxidants also find their way in the wastewater. The conventional wastewater treatment plants are unable to degrade these toxic and complex compounds as these are basically designed to transform the pollutants from one phase to another phase. These pollutants thus can pass into groundwater and pose a serious threat to the environment. One of the best ways is breaking and degrading these complex compounds into their smaller by-products. Irradiation of wastewater by electron beams has been observed to provide a mechanism which successfully fulfills this requirement.

In a common scenario, the EB treatment process involves irradiation of water molecules with a high-power EB which causes radiolysis. In this process, water molecules are either excited or ionized by incident electrons until it dissipates nearly all its energy in water. This process results in the cascade formation of free reactive radical species of energy around 50 eV within $10^{-12}$ seconds of irradiation (Le Caer (2011)). These reactive species include both strongly reducing (hydrated electron: $e_{aq}^-$, hydrogen radical: $H^*$) and oxidising agents (hydroxyl radical: $OH^*$) along with other less reactive species like $H^+$, $H_2O_2$, and $H_2$ (Siwek, Edgecock (2020)). Therefore, this treatment falls under the class of Advanced Oxidation and Reduction process (AORP). All the free radical species chemically rearrange themselves by releasing thermal energy and are termed as spurs. The concentration (the number of moles) of spurs is determined by the radio-chemical yield known as G-Value (shown in bracket below).

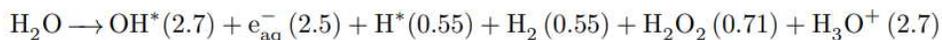

$$H_2O \longrightarrow OH^*(2.7) + e_{aq}^-(2.5) + H^*(0.55) + H_2(0.55) + H_2O_2(0.71) + H_3O^+(2.7)$$

The OH radical is one of the most potent oxidising agents, reacting with many organic impurities with high reaction rate. It mineralises, decomposes, polymerizes, and detoxifies the impurities in wastewater (Capodaglio (2017); Siwek, Edgecock (2020)). Studies have shown that electron irradiation decomposes complex molecular contaminants which are non-biodegradable into smaller molecular structures which then become biodegradable. Electron beams have therefore been used in various wastewater treatment facilities in different countries such as Austria, Brazil, Iran, South Korea, and recently in China. Most common use of EB irradiation in these treatment facilities are chemical reduction and degradation of dyes, phenols, halomethanes and other complex and toxic non-biodegradable compounds (Hossain Kaizar (2018)). Moreover, studies have also been performed on textile industry and paper mill effluent with the combination of EB irradiation with existing conventional techniques (Han et al. (2012)). The results show that EB irradiation can be more efficient while combining with the conventional treatment. Furthermore, these existing EB treatment plants have pointed out that the savings resulting from installing an EB system in a typical setup would offset the expenditure of setting it up in just five years (Shin et al. (2002)). Apart from this, electron beam sources i.e. electron accelerators have been developed with high power efficiency bringing down the cost and treatment time.

Prior studies on pharmaceutical effluent have been conducted using a simulated sample while in this paper, the effect of electron beam (EB) irradiation on real pharmaceutical wastewater is studied. For this study, the samples were collected from different stages of the treatment process of a pharmaceutical industry. The present paper is an effort to optimize the EB irradiation in conjunction with conventional methods like air flow, ozonation, coagulation, and air flow with $TiO_2$ to increase the efficiency and reduce the treatment cost. The main goal was to reduce the organic content of wastewater effectively in a sequential treatment procedure. To obtain optimum removal effectiveness using E-beam treatment, the impacts of operating factors such as absorbed dose and coagulant material concentration were investigated. The degradation of all impurities in wastewater by electron beams are not discussed here. Only COD has been used to quantify the effect of EB irradiation and combinations on water quality. As this is a simple, quick analysis with high accuracy and it covers a wide range of chemical compounds in wastewater.

## 2 Materials & methods

To investigate the efficacy of the EBT in real life situations, samples from different stages of a wastewater treatment plant of a pharmaceutical industry were collected. As mentioned earlier, the EBT works better in conjunction with conventional methods. Depending on the type of effluent, sometimes EBT works better if it precedes the conventional treatment and sometimes if it follows the conventional treatment. In order to investigate whether this order affects the final outcome in our case, we have chosen three samples from different stages of their treatment plant (as outlined in section 2.1).

### 2.1 Wastewater sample

Samples for the present study were collected from a pharmaceutical industry that contains heavy and sizable molecular compounds, primarily organic chemicals, heavy aromatic compounds such as phenol and its derivatives, acids, esters, and heterocyclic compounds. The industry has its own wastewater purification scheme with two different streams. Each stream has different stages of treatment as shown in figure 1 and 2. The treatment process includes soil bio-treatment (SBT), Fenton process, reverse osmosis (RO), multiple-effect evaporator (MEE), and stripper column. Three samples were selected for the study in this paper. The first sample was taken from stream-1 feed to stripper line, second is from stream-1 feed to RO, and the third one is taken from stream-2 after the Fenton process and are named as S1, S2, and S3 respectively.

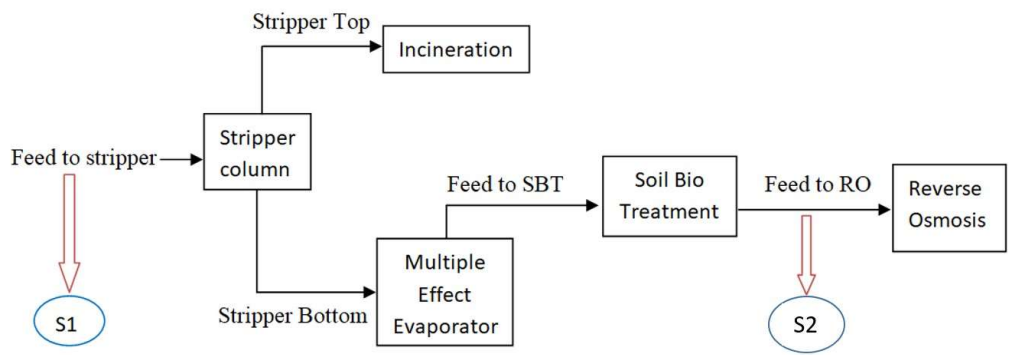

Figure 1: Pharmaceutical wastewater stream-1 and its treatment stages

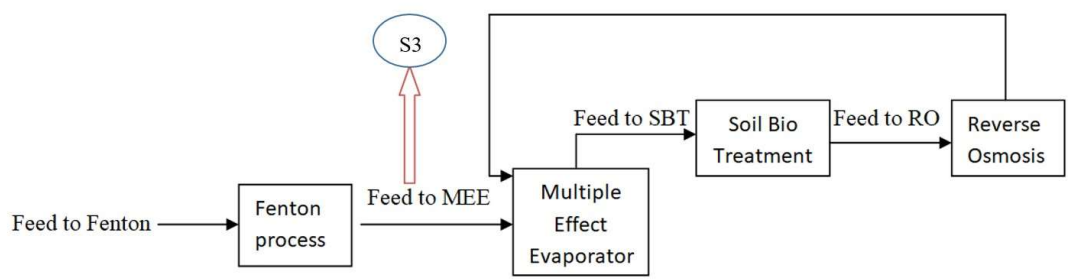

Figure 2: Pharmaceutical wastewater stream-2 and its treatment stages

The sample S1 is a raw wastewater from the first stream, it contains organic polyaromatic compounds like toluene, phenols, nitrophenols, nitroaniline, pharmaceutical compounds and their intermediates with solvents like hexane, xylene, aqueous ammonia etc (Gadipelly et al. (2014)). The chemical and biological materials in form of suspended solid (metals and insoluble compounds) and liquids also persist in this effluent.
Sample S2 is taken after the Soil Biotreatment (SBT) process which is an environment friendly wastewater treatment process which uses microorganism (Indigenous soil microflora) and minerals to convert organic impurities into $CO_2$ in wastewater. This process also includes adsorption and filtration, so the suspended and dissolved solids are removed by adsorption and biodegradation and uptake by green plants. This plant also has limitations over non-biodegradable impurities like other biological processes. The influent water in SBT process has comparatively lower amount of COD values (i.e., 1315 mg/L). The COD reduction by this process was observed around 10%, which means that most of the remaining COD was because of non-biodegradable compounds. Phenol is one of the non-biodegradable compounds still present with concentration of 18.75 ppm in the effluent from SBT.

The Sample S3 is collected after the Fenton process which is an advanced oxidation process in which hydroxyl radicals (*OH) are produced by chemical reaction of hydrogen peroxide ($H_2O_2$) with ferrous ions ($Fe^{2+}$). It has been published that about 95% of COD can be removed from pharmaceutical wastewater which contain chloramphenicol, paracetamol by this process (Gadipelly et al. (2014)). In this process, the hydroxyl radicals oxidize and reduce the toxicity by breaking most of the non-biodegradable compounds into biodegradable compounds. Influents into this process comes from the second wastewater stream of the pharmaceutical industry with COD value of 20000 mg/L. This second stream also contains effluents from dyes and pigment manufacturing units. The COD removal of this process depends on the $FeSO_4$ (ferrous ion) dose. However, after a concentration of 500 mg/L the COD cannot be further removed (Aljuboury et al. (2014)). Because of high concentration of impurities, the Fenton process could not remove all dyes and pharmaceutical impurities from the second wastewater stream (Xu et al. (2004); Nidheesh et al. (2013)). All of these samples were analysed before the irradiation. Effluent quality parameters of the samples are given in Table 1.

Table 1: The initial parameters of raw samples from different stages of the pharmaceutical industry

| Effluents | pH | COD (ppm) | TDS (ppm) |
|---|---|---|---|
| Feed to stripper (S1) | 11.11 | 19,820 | 853 |
| Feed to RO (S2) | 7.74 | 704 | 1,421 |
| MEE Feed Water (S3) | 7.52 | 2,336 | 13,165 |

The pH and conductivity of samples were monitored by OAKTON PH 150 pH meter and Hach Sension 156 multimeter, respectively. COD was analysed using the standard titration method (closed reflux, titrimetric method) as described in the APHA manual (APHA et al. (1981)) with HANNA USA ($10^o$C to $160^o$C) COD Digester.

2.2 Materials

All the solutions that were needed for the COD estimation, as per the APHA manual, were prepared in deionized water (APHA et al. (1981)). Coagulation before and after irradiation was performed with Ferric sulphate ($Fe_2(SO_4)_3$) as coagulant material. The concentration of coagulant material was used up to 32mg/L. The sample used in this research contains high amount of ammonium nitrogen (2640 ppm) which on irradiation results in the final product of nitrate and nitrite compounds. One of the possible reactions is given below (Son et al. (2013)).

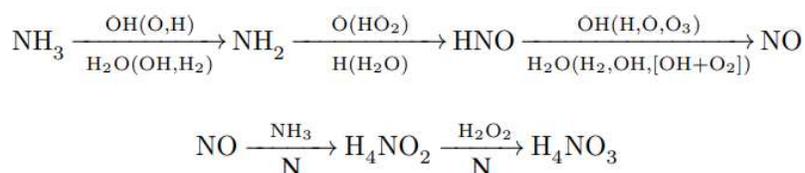

$$NH_3 \xrightarrow[H_2O(OH,H_2)]{OH(O,H)} NH_2 \xrightarrow[H(H_2O)]{O(HO_2)} HNO \xrightarrow[H_2O(H_2,OH,[OH+O_2])]{OH(H,O,O_3)} NO$$

$$NO \xrightarrow[N]{NH_3} H_4NO_2 \xrightarrow[N]{H_2O_2} H_4NO_3$$

Sulphamic acid was used for nitrite free COD estimation as it was observed that nitrite interferes with the estimation of COD (Ferraz, Yuan (2017)). Sulphamic acid concentration required is 10 times the nitrite amount in sample as per the APHA manual (APHA et al. (1981)). Nitrite concentration considered was 200 mg/l of the sample as an initial guess with which concentration of sulphamic acid used in testing is 2 mg/ml of the sample. Higher concentration of sulphamic acid of 4 mg/ml was also used to check the interference. The results are discussed in the results and discussion section.

2.3 Experimental setup

Experimental study to optimise the treatment efficacy of pharmaceutical effluents by combining electron beam irradiation with conventional techniques

Irradiation of effluent samples was performed using a 6 MeV, S-band standing wave side coupled, normal conducting, pulsed linac, situated at Society of Applied Microwave Electronics Engineering and Research (SAMEER) in IIT Bombay campus. The average electron beam current during experiments was 16 $\mu A$.

In order to optimize the dose deposition in the effluent, a detailed Fluka-Flair simulations were carried out (Battistoni et al. (2007)). Samples were irradiated in a 100 mL standard cylindrical borosilicate glass beaker having a diameter of 46.2 mm which is the same as that used in the simulation. The penetration depth of 6 MeV electrons in pure water is 32 mm. The sample volume thus used was 55 ml resulting in a water layer of 33 mm thickness.

**Alignment:** The accelerator is basically designed for cancer treatment which delivers a pencil beam of radius 6 mm. Due to the pencil beam the dose deposition was not uniform as shown in figure 3. The sample was thus continuously stirred using a magnetic stirrer for dose uniformity. As shown in figure 4, the beaker was placed beneath the exit window of the accelerator.

2.4 Dose calculation

Fluka-Flair simulations were carried out with the same geometry as that during the experiments to determine the absorbed dose in the sample. Dose associated with the irradiation time is given in the Table 2.

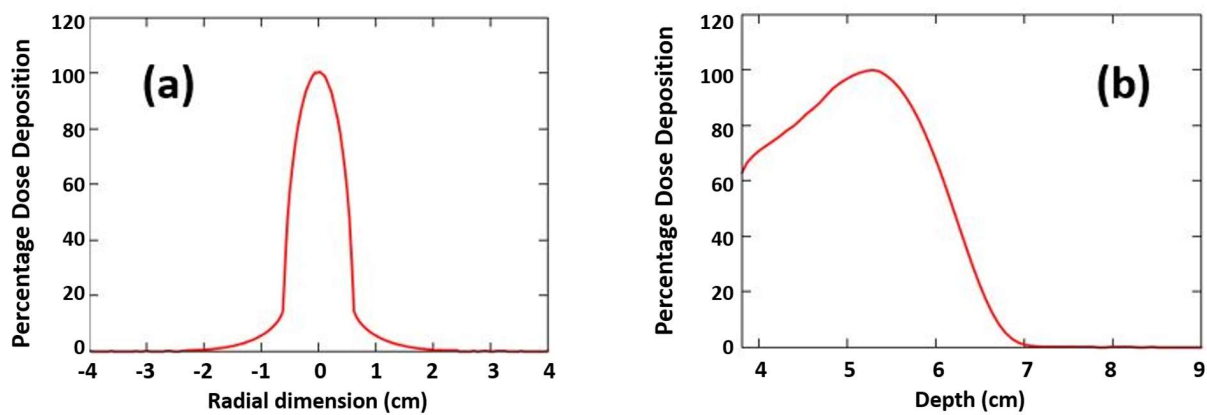

Figure 3: Fluka-Flair simulations for (a) Dose deposition along radial direction of electron beam, (b) and Depth dose profile in water sample

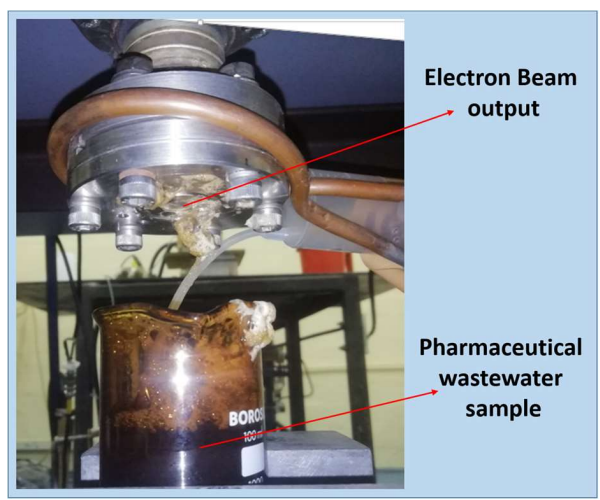

Figure 4: Experimental setup of electron beam accelerator with wastewater

## 3 Results and Discussion

The results of irradiation on three pharmaceutical effluent samples are presented here in terms of COD which is reported as a function of absorbed dose. Samples were also treated with a combination of processes such as irradiation + airflow (EBA), irradiation + ozone-flow (EBO), irradiation + $TiO_2$ + airflow (EBTA), and coagulation before (CEB) and after irradiation (EBC) to check the improvement in water quality parameters after irradiation. The behaviour of treatment of all three effluent samples by different processes were observed to be different. Therefore, the results are presented and discussed separately sample wise.

### 3.1 S1 Sample (Raw effluent)

The irradiation of this waste water by electron beam generates the most reactive specie i.e., hydroxyl radicals (OH*). OH radicals have a strong electrophilicity towards high electron density molecules such as the phenyl rings. It has also been established that pharmaceutical compounds like carbamazepine, ketoprofen, Mefenamic acid, clofibric acid (2-(4-Chlorophenoxy)-2-methylpropanoic acid), diclofenac and phenol are attacked by OH radicals with rate constant of the order of $10^9$ (mol$^{-1}$dm$^3$ s$^{-1}$) (Kimura et al. (2012)). The multiclass pharmaceutical residue in wastewater like Atenolol, Atorvastatin, Azithromycin, Caffeine, Carbamazepine, Ciprofloxacin, Clarithromycin, Diclofenac, Erythromycin, Ibuprofen, Ketoprofen, Losartan, Metoprolol, Naproxen, Sulfamethoxazole, Trimethoprim, Valsartan and Xylazine are also degraded by electron beam irradiation with radiation dose ranging from 0.5 to 25 kGy (Reinholds (2017)).

Table 2: Dose associated with irradiation time

| Irradiation time (s) | 15 | 30 | 60 | 120 | 480 |
|---|---|---|---|---|---|
| Associated Dose(kGy) | 25 | 50 | 100 | 200 | 800 |

The rate constant of OH radicals for many organic compounds are around $10^6$ to $10^{10}$ mol$^{-1}$dm$^3$ s$^{-1}$ (Buxton et al. (1988)). Volatile organic compounds like toluene, ethylbenzene, xylenes and chlorobenzene are easily degraded by the electron beam process (Do-Hung Han (2003)). Moreover, the detergent compound can also be easily degraded by the radiation dose, in the range of 3kGy to 100 kGy. (Selambakkannu et al. (2021)). The above data clearly shows that electron beam irradiation can degrade most of the impurities in pharmaceutical wastewater. It has been found that radiation treatment with conventional techniques (coagulation and biological process) can reduce the COD value by 89 to 94 % in pharmaceutical wastewater at a radiation dose from 25 to 100 KGy (Rahil Changotra (2019)). With this objective, the irradiation of raw wastewater by electron beam was carried out. After irradiation with different doses, suspended solids were formed and a coagulation process was required to remove those suspended solids. The suspended solids formation may have different reasons, one could be the conversion of organic compounds into insoluble compounds at higher dose (Al-Ani, Al-Khalidy (2006)) and other is the reduction of dissolved metals in wastewater into their insoluble form (Unob et al. (2003)). As discussed above the OH radicals prefer to attack the ortho carbon atom in the benzene ring for phenol-like compounds. For more than one benzene ring compound like naphthalene, the OH radical attacks on the para-position carbon atom (Zhengcheng Wen, Xu (2021)). The broken aromatic benzene ring results in long chain aliphatic compounds. Further possible reactions of these intermediate compounds with water radiolysis products are shown in figure (5). The intermediate products can be transformed into carboxylic acid and esters in the presence of oxygen molecules (Emmi, Takács (2008)). Large chain lengths of carboxylic acid decrease the solubility in water. The larger ester group may lead to polymerisation process in wastewater during irradiation. We have also found that during airflow with electron beam irradiation, a significant number of flocs were observed after irradiation.

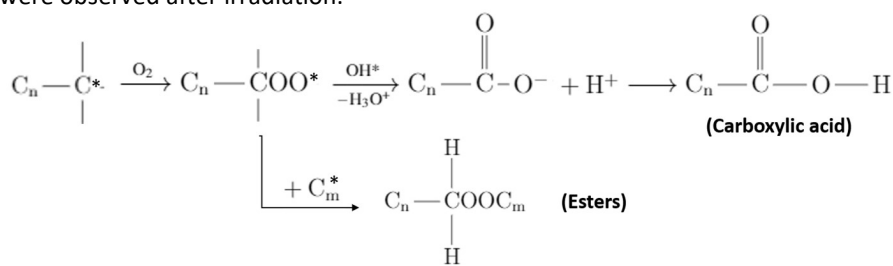

Figure 5: Possible combination of water radicals with intermediate compounds Emmi, Takács (2008)

Figure 6 shows the percentage reduction in COD values of S1 sample after irradiation (EB), irradiation followed by coagulation (EBC) and coagulation followed by irradiation (CEB). When the sample was irradiated by EB, the percentage COD reduction was 11.6% for 25 kGy dose and reduction was 0.3% for 100 kGy. This rise in COD value with irradiation time can be attributed to conversion of complex non-biodegradable compounds into two or more simple biodegradable compounds (Bae et al. (1999); He et al. (2015); Siwek, Edgecock (2020)). It was also observed that the turbidity of water increased after irradiation possibly due to precipitation of soluble impurities. Further removal of the impurities was done with the help of coagulation by ferric sulphate. The maximum reduction in COD was observed at 100 kGy i.e. 33.87 % with EBC and 43.95% with CEB. The amount of coagulant used is incorporated in the name of the corresponding sample. As the presence of nitrite can interfere with COD measurements. Hence, we introduced a sulphamic acid concentration of 2 mg/ml and 4 mg/ml and measured COD values with EB, CEB and EBC for 25 and 100 kGy. The measurement with and without sulphamic acid is shown in the figure 6 with the error bar (±1.62 %).

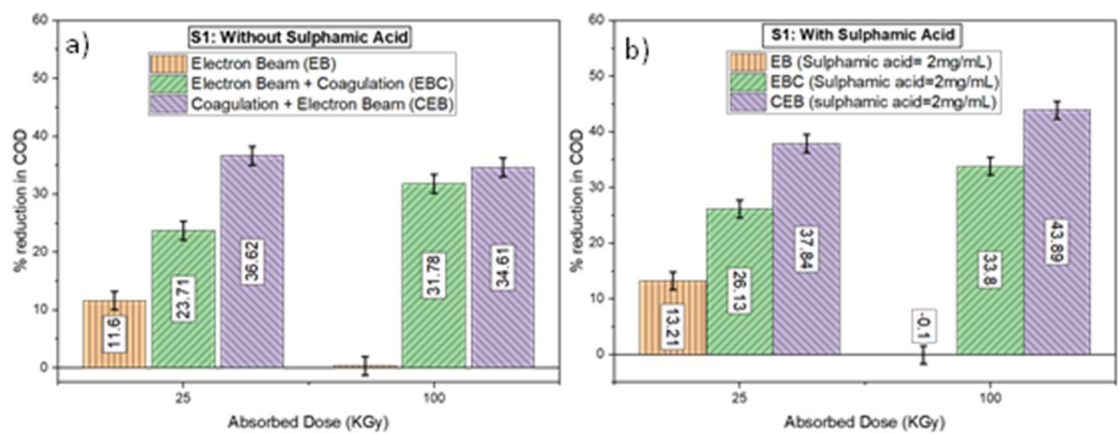

Figure 6: Percent COD reduction after treatment by EB, EBC and CEB process for different irradiation time for S1 with and without sulphamic acid

3.2   S2 Sample (After Soil Bio-treatment)

This sample contains smaller structures of non-biodegradable compounds which could not be treated by the SBT plant. The EB process can easily break these impurities into smaller structures or fewer ring structures (Silva (2016); Tominaga et al. (2018); Qi Zou (2021)). This would be the main reason for the initial COD increment in figure 7 (a) with EB while on combining with coagulation the percentage COD reduction increases to 65.22% at 200 kGy. In this sample, coagulation was not observed before irradiation (CEB). Therefore, the data for CEB is not presented for this sample. This sample was irradiated by EB for 25, 100, and 200 kGy. COD reduction was 65% for EBC processes with 200 kGy which is far better than 34% of reduction with EB after 800 kGy where 800 kGy dose was given just to check the maximum reduction possible. COD reduction observed after removal of nitrite interference by addition of sulphamic acid with concentration of 2 mg/ml is represented by figure 7(b). The results showed that the maximum COD reduction was 54% for 800 kGy while with the EBC process the maximum reduction was 66.9% after 200 kGy which clearly indicates that the EBC process is very effective for this sample.

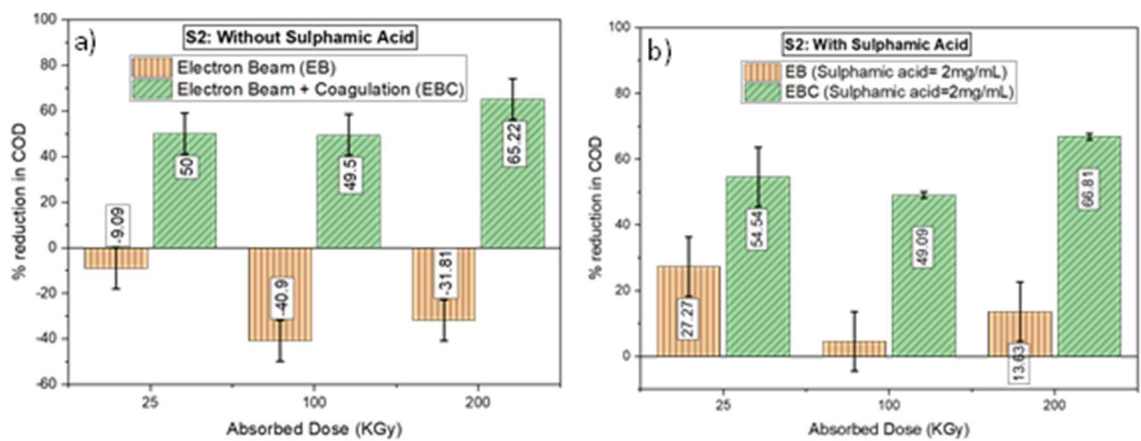

Figure 7: Percent COD reduction after treatment by EB and EBC process for different irradiation time for S2 with and without sulphamic acid

### 3.2.1 The Scavengers Effect:

The presence of compounds like nitrate ($NO_3^-$), nitrite ($NO_2^-$), oxygen molecule ($O_2$), carbonate (CO), and bicarbonate ($HCO_3^-$) act as radical scavengers (Peter, Eschweiler (1999)). These scavengers react with free radicals before the radicals get a chance to act on impurities (Pikaev et al. (1997); M.H.O et al. (2001); Emmi, Takács (2008)). However, depending on the type of impurities these scavengers can be used advantageously to annul either reducing or oxidising radicals. In these experiments, three different radical scavengers namely air, $TiO_2$ powder, and ozone were used. These scavengers scavenge the reducing agents and convert them into oxidizing species, which further help in mineralising the impurities. For example, with air flow, the oxygen molecules from air scavenge reducing species ($e^-_{aq}$) and convert it into oxidising agents ($O_2^-$ and $HO_2^-$). The $TiO_2$ powder is a semiconductor with a band gap energy of 3.2 eV. When electrons with energy higher than 3.2 eV interact with $TiO_2$ molecules, an electron-hole pair is produced, and this electron-hole pair also results in the formation of new oxidising agents ($OH_{abs}$, in the presence of oxygen molecules ($O_2^-$, OH*). The ozone gas converts reducing agents (aqueous electron and hydrogen atom) into $O^*_3$ and $HO^*_3$. The combination of EB irradiation with different treatment processes like airflow, ozone flow, and $TiO_2$ powder with airflow was performed with the hope to see further reduction in COD. However, the COD reduction was observed only in the S2 sample as shown in figure 8. The percentage reduction in COD value observed in the experiment is between 27 to 30% with error bar of ±4.5%.

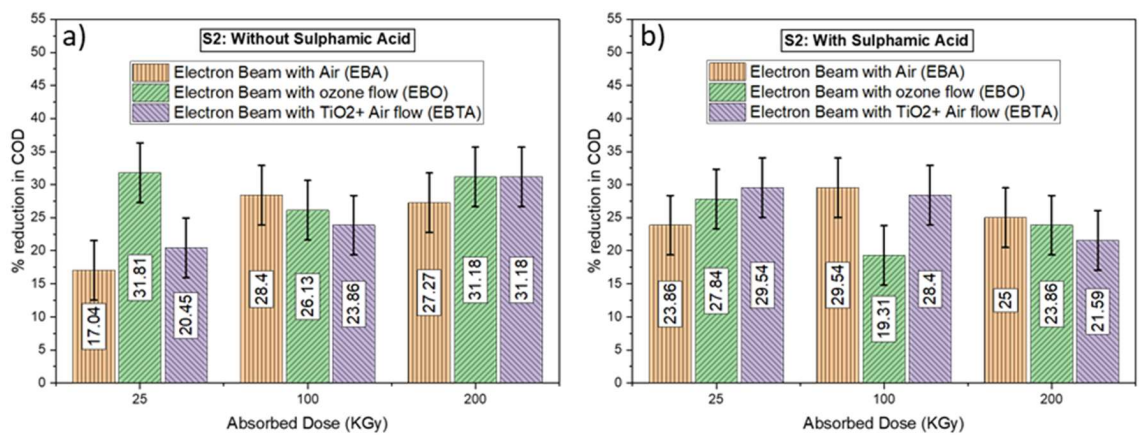

Figure 8: The percent reduction in COD for S2 wastewater after irradiation with airflow (EBA), ozone (EBO) flow and in the presence of $TiO_2$ powder with airflow (EBTA)

### 3.3  S3 Sample (After Fenton process)

This sample is the effluent from the Fenton process and after filtration, the color of this sample was blood red which indicates that low amounts of dyes and organic pollutants were present in this sample. Since the water is already filtered, the suspended solids were not detected. There was no effect of coagulation by ferric sulphate before and after irradiation. Fenton process is an advanced oxidation process which can degrade aromatic compounds like phenol up to 82% (Kavitha, Palanivelu (2004)). Thus, the irradiation of this wastewater sample leads to complete mineralization of the by-products as shown in the figure 9 along with remaining dyes and other remaining products (Chengji Zhang (2021)). This might be the main reason for increase in COD reduction as the dose increases. This sample was irradiated by EB for 25, 100, 200 and 800 kGy. The maximum COD reduction of 54.7% was observed for 8 minutes of irradiation without sulphamic acid as shown in figure 10 whereas the reduction was 58.9% with sulphamic acid.

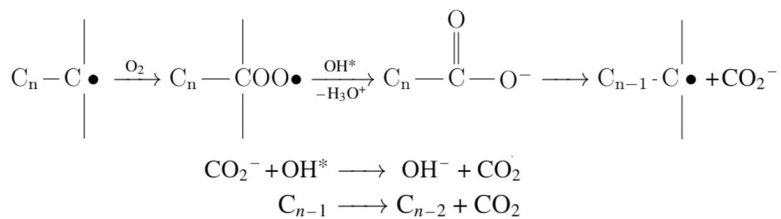

$$C_n - \overset{|}{\underset{|}{C}} \bullet \xrightarrow{O_2} C_n - \overset{|}{\underset{|}{C}} - COO\bullet \xrightarrow[-H_3O^+]{OH^*} C_n - \overset{O}{\underset{|}{\overset{\|}{C}}} - O^- \longrightarrow C_{n-1} - \overset{|}{\underset{|}{C}}\bullet + CO_2^-$$

$$CO_2^- + OH^* \longrightarrow OH^- + CO_2$$
$$C_{n-1} \longrightarrow C_{n-2} + CO_2$$

Figure 9: Carbon chain mineralization process by water radiolysis products in the presence of oxygen molecules.

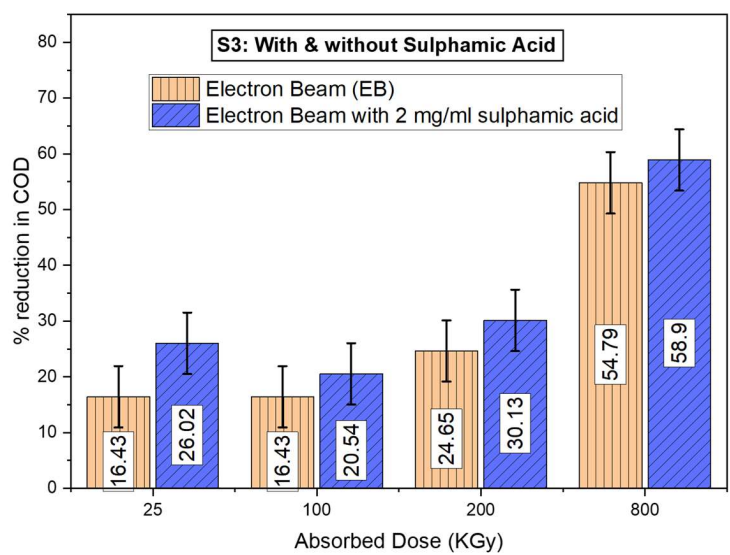

Figure 10: The variation in COD with respect to irradiation time for S3 wastewater sample

## 4  Conclusion

In this study, we have irradiated three pharmaceutical wastewater samples with the help of a 6 MeV electron accelerator. The study was to optimize the EB treatment procedure in conjugation with conventional treatment methods like coagulation. Sample S1 was irradiated by EB and maximum percentage COD reduction was found to be 11.6% for 25 kGy. Treatment of S1 with EBC and CEB gives a maximum reduction of 33.87% and 43.95% respectively, in COD value for 100 kGy. The obtained results suggest that CEB is more effective in this case. Whereas, the reduction of 65% was obtained by EBC, just after 200 kGy in sample S2. Results obtained for sample S1 and S2 clearly demonstrate that the conjugation of EB treatment with coagulation is more efficient than EB treatment alone. For sample S3, the maximum COD reduction of 54.7% was observed for 800 kGy. Along with COD reduction, we have also observed precipitate settled for the irradiated samples which shows that soluble impurities are converted into insoluble by the application of EB treatment. This study suggests that such difficult effluents can be treated by EB irradiation by combining it with a suitable conventional treatment method. Further optimization can help to fit the EB process in the existing facility of an industrial wastewater treatment plant.

## Compliance with Ethical Standards:

Disclosure of potential conflicts of interest: The authors declare that they have no known competing financial interests or personal relationships that could have appeared to influence the work reported in this paper.
Research involving Human Participants and/or Animals: Not applicable
Informed consent: Not applicable

## Statements and Declarations

Funding: The authors declare that no funds, grants, or other support were received during the preparation of this manuscript.
Author Contributions: Pankaj Kumar: Conceptualization, Material preparation and analysis, Writing - original draft. Raghava Varma: Supervision, Writing - review & editing. Abhishek Pathak: Writing - review & editing. Manisha Meena, Anjali Bhagwan Kavar and Pragya Nama: Data collection and analysis. Abhay Deshpande, Tanuja Dixit, Ramamoorthy Krishnan and Chandrakant Nainwad: Experimental setup design and irradiation. All authors commented on previous versions of the manuscript. All authors read and approved the final manuscript.
Research Data Policy: The data that support the findings of this study are available from the corresponding author, [Pankaj Kumar], on special request.

## Acknowledgements

We would like to thanks Prof. Sanjeev Chaudhari (Centre for Environmental Science & Engineering, IIT Bombay) and Prof. Arindam Sarkar (Chemical Engineering, IIT Bombay) for their valuable support and assistance at every stage of this work. Raghava Varma is indebted to Prof. Atsuto Suzuki for inspiring and motivating him to attempt something different for societal benefits.